\documentclass[twocolumn,prb,aps]{revtex4}
\usepackage{amsmath}
\usepackage{amssymb}
\usepackage{graphicx}
\usepackage{dcolumn}
\usepackage{bm}

\begin{document}

\preprint{Phys.Rev.B}

\title{Transition from insulating to metallic phase induced by in-plane
magnetic field in HgTe quantum wells}

\author{G. M. Gusev,$^1$ E. B. Olshanetsky,$^2$ Z. D. Kvon,$^{2,3}$
O. E. Raichev,$^4$ N. N. Mikhailov,$^2$ and S. A. Dvoretsky,$^{2}$}

\affiliation{$^1$Instituto de F\'{\i}sica da Universidade de S\~ao
Paulo, 135960-170, S\~ao Paulo, SP, Brazil}
\affiliation{$^2$Institute of Semiconductor Physics, Novosibirsk
630090, Russia}
\affiliation{$^3$Novosibirsk State University, Novosibirsk, 630090, Russia}
\affiliation{$^4$Institute of Semiconductor Physics, NAS of Ukraine, Prospekt Nauki 41, 03028,
Kiev, Ukraine}

\date{\today}

\begin{abstract}

We report transport measurements in HgTe-based quantum wells with
well width of 8 nm, corresponding to quantum spin Hall state, subject
to in-plane magnetic field. In the absence of the magnetic field the
local and nonlocal resistances behave very similar, which confirms the
edge state transport in our system. In the magnetic field, we observe
a monotonic decrease of the resistance with saturation of local resistance,
while nonlocal resistance disappears completely, independent of the
gate voltage. We believe that these evidences of metallic behavior
indicate a transition to a gapless two-dimensional phase, according
to theoretical predictions. The influence of disorder on resistivity
properties of HgTe quantum wells under in-plane magnetic field is
discussed.


\end{abstract}

\maketitle

\section{Introduction}

Recent years have witnessed an astonishing growth in research on topological
insulators, the materials that have a bulk band gap like an ordinary insulator
but support conducting states on their edge or surface \cite{hasan, qi, moore,
moore2, kane, bernevig, bernevig2}. The two-dimensional (2D) topological insulators
can be classified either into quantum Hall effect (QHE) state or quantum spin Hall
effect (QSHE) state. The edge modes in the traditional integer quantum Hall system
are chiral because time reversal symmetry is broken by the magnetic field.
In contrast, QSHE state is described by pairs of counter-propagating edge modes
with opposite spin polarizations (Kramers pairs) so that the time reversal
symmetry is maintained. Experimentally measured 4-probe resistance in a
micrometer-sized Hall bar fabricated from HgTe/CdHgTe quantum well structure
demonstrated a quantized plateau $R_{xx} \simeq h/2e^{2}$ in the absence of a
perpendicular magnetic field \cite{konig}. This fact, supported by a theoretical
consideration of the possibility of QSHE regime in HgTe quantum wells \cite{bernevig2},
has been taken as a definitive evidence for QSHE state. One more experimental evidence
for QSHE is a nonlocal transport \cite{roth}, when application of the current between
a pair of contacts creates a net current along the sample edge and causes a
voltage at any other pair of the contacts. In QHE regime, when the edge modes
are chiral, the nonlocal response requires backscattering between opposite edges,
which occurs via the bulk states and disappears at purely integer filling
of Landau levels. In QSHE state, the nonlocal response always exists because of the
presence of two counter-propagating edge modes at the same edge. Moreover, the
spin-flip scattering between these modes, which is important in HgTe
quantum well structures of several micrometer size \cite{konig}, makes the
nonlocal resistance considerably larger than the resistance quantum \cite{gusev}.

A remarkable property of HgTe quantum wells is the opportunity to create
different band structures for 2D electrons. The ordinary 2D insulator state is realized
at small well widths (approximately, up to 6.3 nm), while 2D topological insulator
(QSHE) state exists at larger well widths. The large-width (20 nm and wider)
quantum wells support semimetallic 2D state \cite{kvon, kvon1, gusev1}
with overlapped conduction and valence 2D bands. Recent theoretical consideration
\cite{raichev} suggests one more possibility, when application of a strong (of the
order 10 T) magnetic field parallel to the plane of HgTe quantum well in QSHE state
rebuilds the subband structure and creates a gapless 2D state with electron energy
spectrum similar to that of bulk HgTe. Preliminary experimental studies \cite{gusev}
indeed show a strong decrease of the resistance of HgTe quantum well structures with
increasing in-plane magnetic field above 10 T, which can be a signature of a transition
to metallic state. Since the problem of this phase transition deserves more attention,
we have carried out additional experiments and theoretical calculations. The nonlocal
resistance measurements are very useful for this purpose, because the appearance of
the metallic state in the bulk of the sample shunts the edge channel transport and
makes the nonlocal resistance exponentially small, practically equal to zero.

In this paper we present results of resistance measurements in 8 nm wide wells of
different sizes in the presence of the in-plane magnetic field. The four-terminal
resistance in samples with gate size $13 \times 7$ $\mu$m$^{2}$ is $R_{xx} \simeq
300$ kOhm and still significantly larger than $h/2e^{2}$. The nonlocal transport
experiments with these devices in the QSHE regime demonstrate that charge transport
occurs through extended edge channels. Experimentally, we show that the in-plane
magnetic field suppresses nonlocal resistance completely, while local transport
demonstrates a monotonic decrease of the resistance with saturation. The evolution
of the 2D subband structure with increasing magnetic field is calculated by means
of numerical solution of eigenstate problem for the 6x6 matrix Kane Hamiltonian
taking into account strain effects in HgTe well. These calculations confirm the
gapless nature of the metallic state induced by the magnetic field. To describe
the behavior of the resistance near the phase transition, we analyze the influence
of smooth inhomogeneities of the system such as variations of the well width and
electrostatic potential.

The paper is organized as follows. In Sec. II we characterize the samples and present
the results of resistance measurements in zero magnetic field. Section III describes
experimental results in parallel magnetic field. A theoretical analysis is presented
in Sec. IV. The final section contains a brief discussion of the results.

\section{Experiment in zero magnetic field}

\begin{figure}[ht]
\includegraphics[width=7cm,clip=]{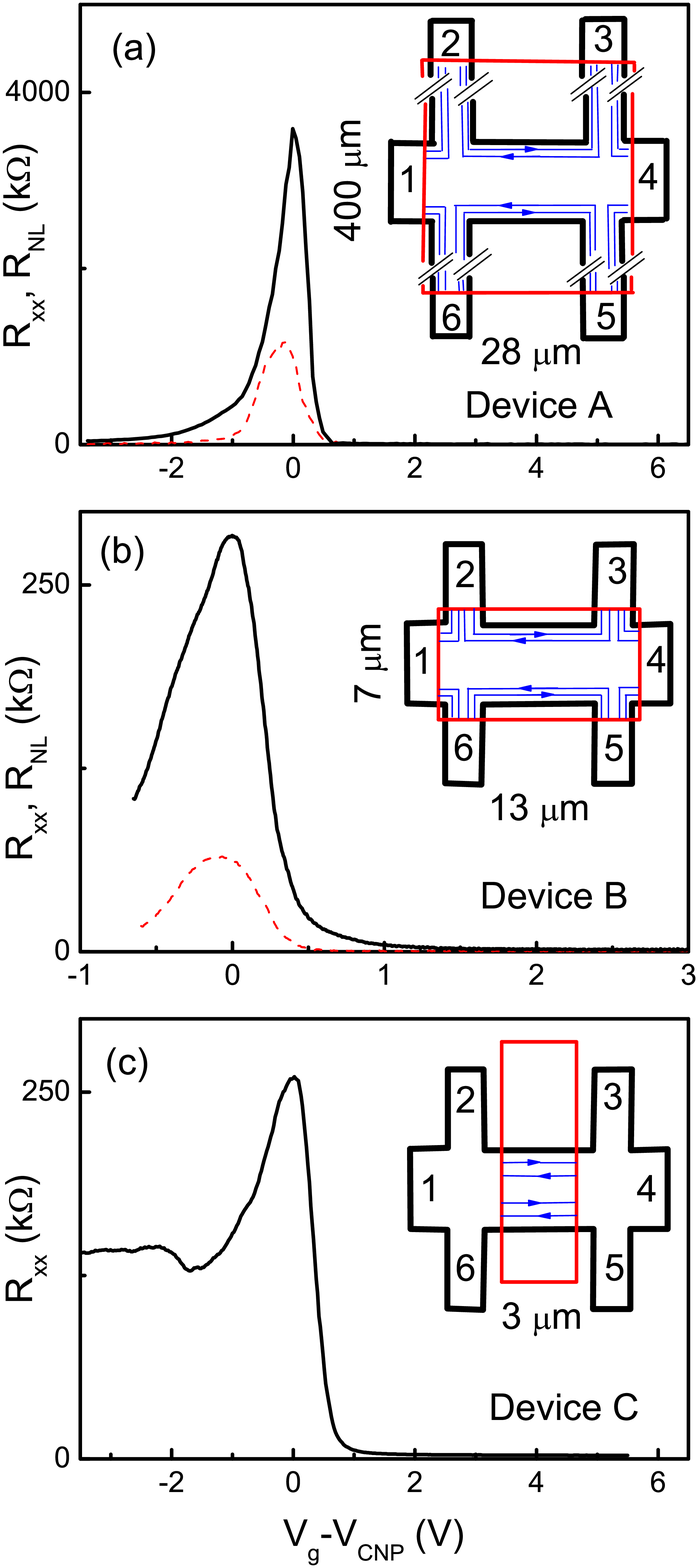}
\caption{\label{fig.1}(Color online) Four-terminal local $R_{I=1,4;V=2,3}$ (black) and nonlocal
$R_{I=6,2;V=5,3}$ (red dashes) resistances as a function of the gate voltage at $T=4.2$ K and $B=0$
for the Hall bar devices A (a), B (b), and C (c) with dimensions of the gate (length and width)
indicated. The numbers indicate the coding of the leads.}
\end{figure}

The Cd$_{0.65}$Hg$_{0.35}$Te/HgTe/Cd$_{0.65}$Hg$_{0.35}$Te quantum
wells with the [013] surface orientations and the widths of 8 - 8.3
nm were prepared by molecular beam epitaxy. The sample consists of
three 5 $\mu$m wide consecutive segments of different length (6.5,
20, 6.5 $\mu$m), and 8 voltage probes. The ohmic contacts to the
two-dimensional gas were formed by the in-burning of indium. To
prepare the gate, a dielectric layer containing 100 nm SiO$_{2}$ and
200 nm Si$_{3}$Ni$_{4}$ was first grown on the structure using the
plasmochemical method. Then, the TiAu gate was deposited. We present
experimental results on three different type of the devices, which
are schematically shown in Fig. 1. Device A (Fig. 1 a) is a
structure with large gate area for identifying nonlocal transport
over macroscopic distances \cite{gusev}. The lengths of the edge
states are determined by the perimeter of the sample part covered by
metallic gate (mostly side branches) rather than by the length of
the bar itself. Devices B and C are structures with small gate area.
Device B is designed for multi-terminal measurements, while device C
has been used for two-terminal measurements. Several devices with
the same configuration have been studied. The density variation with
gate voltage was $1.09 \times 10^{11}$ cm$^{-2}$V$^{-1}$. The
electron  mobility exhibits nonmonotonic dependence on the carrier
density with distinct maximum $\mu_{n} =250\times10^{3}$ cm$^{-2}$/V s
at $n_{s}=2\times10^{11}$ cm$^{-2}$, and hole mobility demonstrates a
saturation $\mu_{p} =20 \times 10^{3}$ cm$^{-2}$/V s with carrier density
$p_{s}=1.5\times10^{11}$ cm$^{-2}$. The magnetotransport measurements
in the described structures were performed in the temperature range
from 1.4 K to 25 K and in magnetic fields up to 12 T using a
standard four-point circuit with a 3-13 Hz ac input of 0.1-10 nA
through the sample, which is sufficiently low to avoid the
overheating effects.

\begin{figure}[ht]
\includegraphics[width=9cm,clip=]{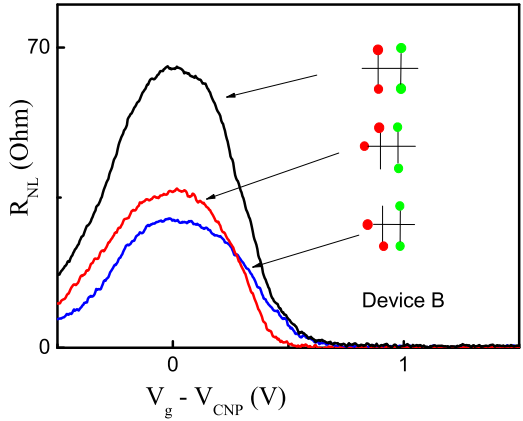}
\caption{\label{fig.2}(Color online) Nonlocal resistance as a
function of the gate voltage for the Hall bar device B and different
nonlocal configurations (from top to bottom): $R_{I=6,2;V=3,5}$,
$R_{I=1,2;V=3,5}$, and $R_{I=1,6;V=3,5}$, at $T=4.2$ K and
$B=0$.}
\end{figure}

The carrier density in HgTe quantum wells can be electrically
manipulated with the gate voltage $V_{g}$. The typical dependence of
the four-terminal local $R_{I=1,4;V=2,3}$ and nonlocal
$R_{I=6,2;V=5,3}$ resistances as a function of $V_{g}$ in device A
(Fig. 1 a) exhibits a sharp peak that is $\sim 10^{4}$ times greater
than the resistance at $V_{g}\sim 2$ V far from the peak position.
The Hall coefficient reverses its sign when $R_{xx}$ approaches its
maximum value \cite{gusev}. This behavior resembles the ambipolar
field effect observed in graphene \cite{sarma}. Thus, the gate
voltage alters the quantum wells from n-type conductor to p-type
conductor via an insulating state. The nonlocal resistance
$R_{I=6,2;V=5,3}$ in the insulating regime has a comparable
amplitude and qualitatively the same position and width of the peak
as the local resistance. Outside of the peak the nonlocal resistance
is negligibly small, as expected for conducting state. Figures 1 b
and 1 c show the experimental data for devices B and C. One can see
that the peak resistance is dramatically reduced in the samples of
few micrometer size, but it is still far higher than the resistance
in the ballistic edge state transport regime. Based on the
conductance of $0.1 e^{2}/h$ for a 3-6 micrometer long sample, one
may conclude that the ballistic regime in our samples is expected if
we reduce the sample length to less than $\sim 0.5$ $\mu$m.  The
understanding of the absence of the quantized transport in
macroscopic samples requires further investigation.

The nonlocal resistance is different in a slightly modified configuration, where the
current passes through the contacts 1 and 6 and the voltage is measured between the
contacts 5 and 3. Figure 2 shows curves for three possible configurations of nonlocal
resistance. The resistance is reduced when current path is shorter, which is
expected for edge state transport with backscattering. Conductivity of 2D topological insulator
is determined by 1D channel (ballistic or diffusive), which connects all contacts
(probes) at the periphery of the sample. Application of the current between any pair
of contacts produces the current circulating along the entire edge. In particular,
for nonlocal configuration $R_{I=6,2;V=5,3}$, which is shown in Fig. 2, two
paths lead from the contact 2 to the contact 6, one (2-3,3-4,5-6) three times
longer than the other (2-1,1-6). The resistance between the contacts can be
substituted by the quantum resistance $R=h/2e^{2}$ in ballistic case or by
\begin{equation}
R = \frac{h}{2e^{2}}(1+\gamma L)
%1
\end{equation}
in diffusive case, where $\gamma^{-1}$ is the mean free path for 1D backscattering
and $L$ is the length of 1D channel between the voltage probes. Within this
approximation one would expect the following ratios between local and nonlocal
resistances: $R_{I=1,4;V=2,3}/R_{I=6,2;V=5,3} \approx 0.22$;
$R_{I=1,4;V=2,3}/R_{I=1,2;V=5,3}\approx 0.11$, which roughly agrees with
experimental data shown in Fig. 2.

\section{Experiment in parallel magnetic field}

Applying a strong magnetic field parallel to a quantum well may lead to several
effects. First, the magnetic field creates an electronic spin polarization,
leading to an increase in spin scattering and an increase in
resistivity \cite{dolgopolov}. Second, the parallel magnetic field
leads to mixing of 2D subband states due to magneto-orbital coupling with the
field, because the wave functions of the confined states have finite widths
determined by the width of the well; see Ref. 17 and, in application to
HgTe wells, Ref. 14.

\begin{figure}[ht]
\includegraphics[width=9cm,clip=]{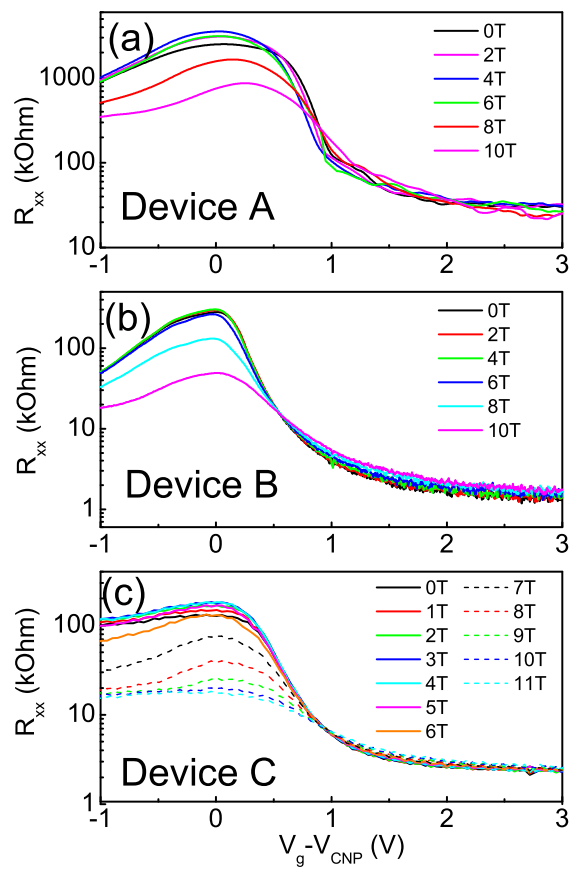}
\caption{\label{fig.3}(Color online) Four-terminal local
$R_{I=1,4;V=2,3}$  resistance as a function of the gate voltage for
the Hall bar devices and for different parallel magnetic fields,
$T=4.2$ K.}
\end{figure}

Figure 3 shows the evolution of the local magnetoresistance with
gate voltage in the presence of the in-plane magnetic field for
three different devices. One can see that the resistance in the peak
in magnetic fields higher than 5 T demonstrates a rapid monotonic
decrease in all samples. This behavior corresponds to a transition
from the insulating state to the gapless metallic state (see detals
in Sec. IV). With increasing gate voltage, the Fermi level is lifted
up into the conduction band so the system is in the metallic state
already at zero magnetic field. The resistance in this case becomes
almost insensitive to the magnetic field, as expected. However, a
weak positive magnetoresistance is present above a critical voltage
in the short devices B and C. Surprisingly, this critical voltage is
independent on the in-plane magnetic field and the corresponding
resistance is of the order of resistance quantum. We have no
explanation of this particular observation.

\begin{figure}[ht]
\includegraphics[width=8cm,clip=]{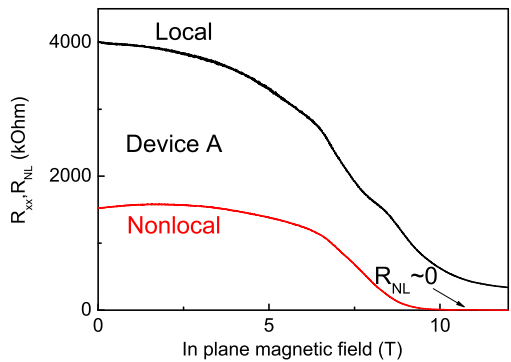}
\caption{\label{fig.4} (Color online) (a) The local
$R_{xx}=R_{I=1,4;V=6,5}$ and nonlocal $R_{I=2,6;V=5,3}$ resistances
as a function of in-plane magnetic field for device A at
$V_{g}=-3.63$ V, T=1.5 K.  The gate voltage corresponds to the
resistance peak. The nonlocal resistance disappears at $B>10$ T.}
\end{figure}

\begin{figure}[ht]
\includegraphics[width=8cm,clip=]{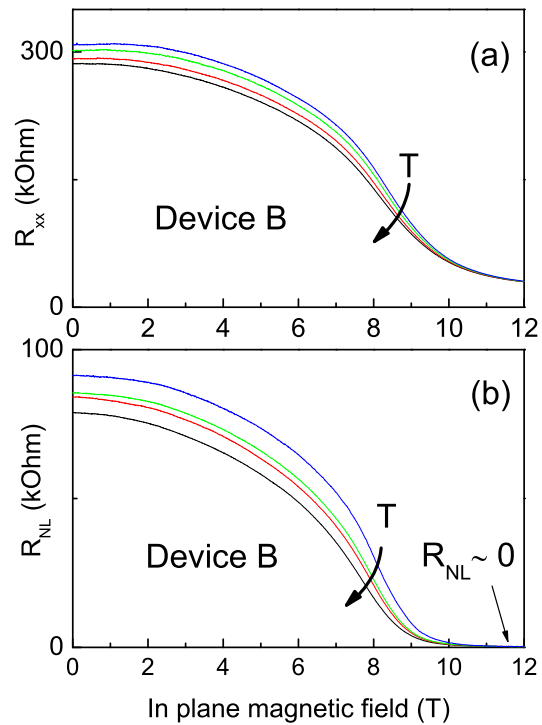}
\caption{\label{fig.5} (Color online) (a) The local resistance
$R_{I=1,2;V=3,5}$ as a function of in-plane magnetic field for
different temperatures $T$ (K): 1.5, 2.5, 3.5, 4.2 at $V_{g}=-2.43$
V. (b) The nonlocal resistance $R_{I=2,6;V=5,3}$ as a function of
in-plane magnetic field for different temperatures $T$ (K): 1.5, 3,
3.6, 4.2 at $V_{g}=-2.4$ V. The gate voltage corresponds to the
resistance peak. The nonlocal resistance disappears at $B=12$ T.}
\end{figure}

In Figs. 4 and  5 we present the magnetic-field dependence of the
local and nonlocal resistances near peak point for different
temperatures for both devices A and B. One can see monotonic
decrease of the resistance with saturation of local resistance,
while nonlocal resistance disappears completely above magnetic field
$B \simeq 12$ T. A rapid decrease of the resistances starts
approximately at $B \simeq 8$ T. In Fig. 5 one can see that the
temperature effect on local resistance is weak.

\begin{figure}[ht]
\includegraphics[width=9cm,clip=]{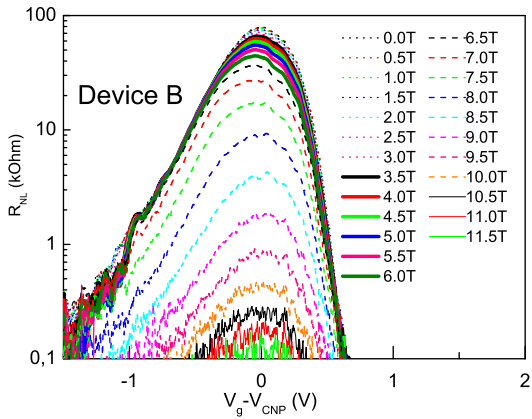}
\caption{\label{fig.6} (Color online) The nonlocal $R_{I=1,2;V=3,5}$
resistance as a function of gate voltage for different magnetic
fields, $T=4.2$ K. }
\end{figure}

The evolution of nonlocal resistance as a function of the gate voltage
with increasing in-plane magnetic field is given in Fig. 6. For the
voltages corresponding to insulating state, we find as much as three
orders of resistance reduction in $B \simeq 12$ T. The resistance peaks
are asymmetric, showing a more rapid decrease with $V_{g}$ in the n-type
region.

Based on our experimental observations we may conclude that the
external parallel magnetic field strongly suppresses local and
completely destroys nonlocal resistance. The nonlocal resistance
could be negligibly small in the presence of a dissipative transport
in the bulk of the sample. Therefore it would be natural to assume
that the in-plane magnetic field produces conducting states in the
bulk and, possibly, suppresses the edge state transport.
Alternatively, we may attribute zero nonlocal resistance to
formation of chiral edge channels similar to the QHE state in
perpendicular magnetic field. Such channels indeed are
dissipationless, and the voltage drop between the contacts is zero.
However, within this scenario, the local resistance would be zero as
well, which disagrees with our observation. Furthermore, we do not
observe any Hall resistance as the in-plane magnetic field
increases.

We explain the observed suppression of nonlocal resistance as a consequence
of a magnetic field-induced phase transition from the insulating (QSHE) phase
to a metallic 2D phase. Below we justify this assumption by a theoretical
analysis of 2D electron spectrum.

\section{Theory}

The phase transition from the insulating to the gapless state has been described
in Ref. 14 by using the effective 2D Hamiltonian derived in the
basis including one interface-like state (e) formed by hybridization of conduction
electrons with light holes in the quantum well and two states corresponding to
the first subbands of heavy holes (h1 and h2). As the effective Hamiltonian
is valid only in a narrow region of 2D electron wave vectors near ${\bf k}=0$,
it is necessary to do more detailed calculations of energy spectrum, which
are not restricted by the limitations of the effective Hamiltonian approach
and are non-perturbative with respect to the magnitude of $B$. To find
the 2D electron spectrum, we have carried out a numerical solution of eigenstate
problem for the 6x6 matrix Kane Hamiltonian which satisfactory describes HgTe/CdHgTe
heterostructures. We also have included the effect of uniaxial strain in HgTe
well due to the lattice mismatch of HgTe and CdHgTe. The material parameters
used for these calculations are given in Ref. 13. The results for
symmetric Cd$_{0.65}$Hg$_{0.35}$Te/HgTe/Cd$_{0.65}$Hg$_{0.35}$Te quantum wells
of width 8 nm grown along [001] direction are shown in Figures 7 and 8. The
energy in these figures is counted from the valence band extremum in bulk HgTe.

\begin{figure}[ht]
\includegraphics[width=9cm,clip=]{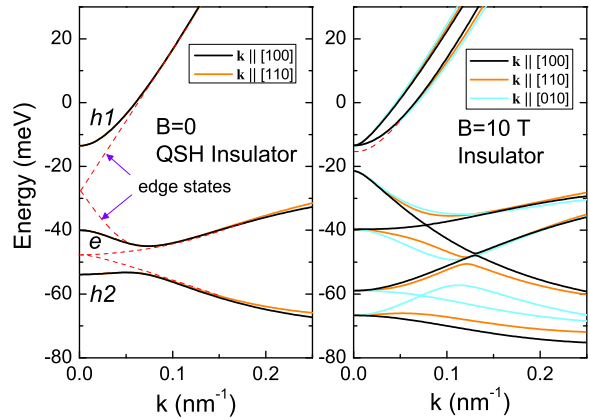}
\caption{\label{fig.7}Dependence of electron energy on 2D wavenumber
$k$ calculated for [001]-grown 8 nm-wide symmetric quantum wells
Cd$_{0.65}$Hg$_{0.35}$Te/HgTe/Cd$_{ 0.65}$Hg$_{0.35}$Te at zero
magnetic field and under in-plane magnetic field $B=10$ T directed
along [100]. Dashed lined schematically show edge state spectrum.}
\end{figure}

Figure 7 demonstrates the effect of magnetic field below the point
of phase transition. Three subbands (e, h1, and h2) are
spin-degenerate at $B=0$, since the well is symmetric. As the
subband structure is inverted (subband e is below subband h1), the
system is in QSHE state. The subband h1 forms the 2D conduction
band, which is almost isotropic. The 2D valence band, formed by e
and h2 subbands, shows a weak anisotropy originating from the
anisotropy of hole states in bulk HgTe as the Luttinger parameters
$\gamma_2$ and $\gamma_3$ are not equal to each other. The
hybridization of e and h2 subbands at finite 2D wave vector $k$
leads to anticrossing of these subbands so that an additional gap is
formed within the 2D valence band. The spectrum of edge states is
shown schematically, based on effective Hamiltonian calculations
\cite{raichev}. As the magnetic field is applied, all the subbands
show spin splitting, so instead of three there are six spectral
branches. The h1 state, whose wave function is symmetric at $k=0$,
remains spin-degenerate at $k=0$. The branches of 2D valence band
have a much higher anisotropy since the spectrum becomes strongly
sensitive to the direction between ${\bf k}$ and ${\bf B}$. In the
field of 10 T, the system is still in the insulating state but the
gap is reduced to approximately 8 meV. The edge states do not
disappear, though their spectrum is considerably modified
\cite{raichev} and is no longer gapless.

\begin{figure}[ht]
\includegraphics[width=9cm,clip=]{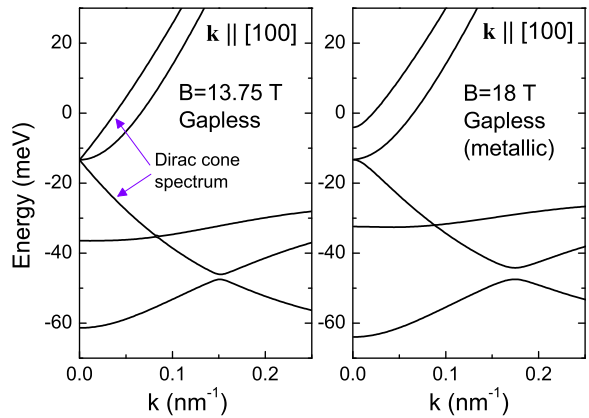}
\caption{\label{fig.8}Dependence of electron energy on 2D wavenumber
$k$ calculated for [001]-grown 8 nm-wide symmetric quantum wells
Cd$_{0.65}$Hg$_{0.35}$Te/HgTe/Cd$_{ 0.65}$Hg$_{0.35}$Te at the
transition field $B=13.75$ T and at $B=18$ T. Both ${\bf B}$ and
${\bf k}$ and directed along [100]. Above the transition field the
spectrum is gapless.}
\end{figure}

Figure 8 shows the spectrum at the transition field, which is equal
approximately 13.8 T for the chosen structure and is almost
insensitive to interface orientation (for [013]-grown wells studied
in our experiment the transition field varies from 13.6 T to 14 T
depending on direction of ${\bf B}$ in the plane), and at a higher
field of 18 T. For clarity, only one direction of ${\bf k}$ is
shown, as for the other directions the picture is qualitatively
similar. The state when the gap disappears is characterized by two
linear branches forming a "Dirac cone" 2D spectrum similar to that
of graphene and a third, parabolic branch passing through the Dirac
point. Further increase in $B$ does not open a gap: the system
remains in the gapless state because the upper spin branch of h1
subband is inverted and acquires a negative effective mass. Above
the point of phase transition, the 2D system is always in metallic
state, without regard to electron density.

The above calculations suggest an abrupt drop of both local and
nonlocal resistance as the magnetic field reaches the transition
field. The experiment, however, shows a smoother decrease of
resistance over a wide region of magnetic fields, and nonlocal
resistance disappears at a field smaller than the calculated
transition field. This is not surprising, because the calculations
are carried out for an ideal quantum well, while realistic quantum
wells are inhomogeneous. The most important kind of spatial disorder
for HgTe wells is the variation of the well width which causes
energy fluctuations of all 2D subbands and variation of the
insulating gap over the 2D plane. Assuming that such variations are
smooth on the quantum length scale, one may use local well width $a$
to calculate the energy spectrum and average the results over the
well width distribution. For the regions with smaller gap the
transition to metallic state occurs at weaker magnetic fields.
Therefore, even when the magnetic field is below the transition
field for a given average well width ${\overline a}$, a part of the
2D plane is already in the gapless state: a number of metallic
clusters (islands) is formed. With increasing field, the average
size of metallic coverage increases. This occurs not only because a
larger fraction of the 2D plane is transferred into the gapless
state (which is always metallic independent of Fermi energy), but
also because the position of the Fermi energy, dictated by the {\em
average} electron density $n$, is shifted out of the gap for some
part of gapped regions of the plane. Assuming a Gaussian well width
distribution $W(a)=\exp[-(a-{\overline a})^2/\delta a^2]/(\sqrt{\pi}
\delta a)$ with $\delta a=1$ nm, we have calculated the fraction of
the insulating coverage of the 2D plane as function of the magnetic
field. The result for charge neutrality point ($n=0$) is presented
in Fig. 9 by the dashed line. Solid lines, plotted also for small
positive and negative $n$, show the results calculated with an
additional kind of disorder, smooth spatial fluctuations of
electrostatic potential which lead to variation of the subband
energies without affecting the gaps between the subbands. This
disorder, usually caused in quantum wells by remote charged
impurities, should be also important in view of negligible screening
of the potential at small electron densities. To bring the
inhomogeneous electrostatic potential $\varphi$ into our
calculations, we again use the Gaussian distribution $W(\varphi)$
with $\delta \varphi=5$ meV.

\begin{figure}[ht]
\includegraphics[width=9cm,clip=]{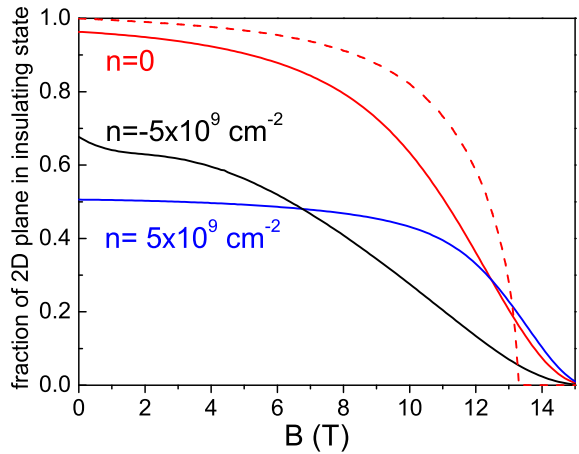}
\caption{\label{fig.9}Magnetic-field dependence of the fraction of
2D plane covered by insulator for spatially-inhomogeneous 8 nm-wide
symmetric quantum wells
Cd$_{0.65}$Hg$_{0.35}$Te/HgTe/Cd$_{0.65}$Hg$_{0.35}$Te (see text for
details). The dashed line shows the result where only well width
variations are taken into account.}
\end{figure}

Figure 9 demonstrates how the insulating coverage of the sample
becomes smaller as the in-plane magnetic field increases. The
presence of inhomogeneous electrostatic potential makes this
behavior smoother but does not affect the picture qualitatively.
Accordingly, the metallic coverage $p$ becomes larger so the
metallic clusters are able to further interconnect to achieve the
percolation process and eventually evolve into a metallic continuum.
A similar physical situation takes place when metallic films are
condensed on an insulating substrate \cite{fahsold}. Even if the
inhomogeneous electrostatic potential is neglected, almost half of
the 2D plane is in metallic state already at 12 T, and the presence
of the electrostatic potential increases the metallic coverage at
this magnetic field. Since different 2D models suggest percolation
threshold around $p=0.5$ (see Ref. 19 and references therein) one
may conclude that the in-plane magnetic field of about 12 T is
indeed capable to induce the phase transition to metallic state
which we detect in our experiment by disappearance of the nonlocal
resistance. We also note an obvious correlation between the
magnetic-field dependence of resistance shown in Figs. 4 and 5
and the behavior shown in Fig. 9.

\begin{figure}[ht]
\includegraphics[width=9cm,clip=]{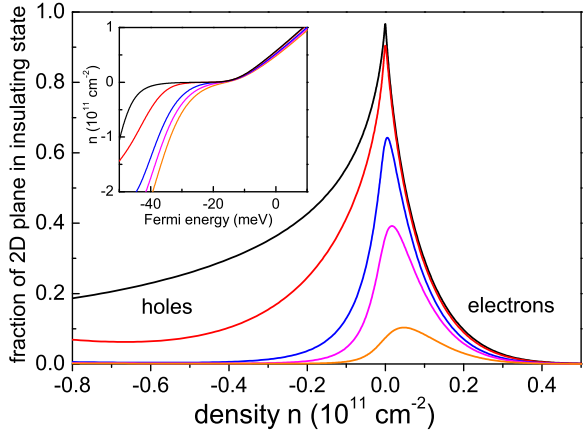}
\caption{\label{fig.10}Fraction of 2D plane covered by insulator as
function of electron density for spatially-inhomogeneous 8 nm-wide
symmetric quantum wells
Cd$_{0.65}$Hg$_{0.35}$Te/HgTe/Cd$_{0.65}$Hg$_{0.35}$Te. The plots
are taken for in-plane magnetic fields $B=0$, 5, 10, 12, and 14 T
(from top to bottom). The inset shows the dependence of electron
density on Fermi energy for the given set of magnetic fields.}
\end{figure}

The fraction of insulating coverage of the sample is highly
sensitive to the electron density which is varied in our experiment
proportional to the gate voltage. Figure 10 shows the density
dependence of the insulating coverage at different magnetic fields,
calculated by taking into account variations of both well width and
electrostatic potential. The inset shows how the electron density
depends on the Fermi energy: for small fields there is a plateau
indicating the presence of the gap, while for higher field this
plateau disappears as the gap is closing. Since the fraction of
insulating coverage correlates with the magnitude of the resistance,
Fig. 10 reproduces the main features of the experimental dependence
of the resistance on the gate voltage (Fig. 6), the maximum at
charge neutrality point and the asymmetry. As follows from our
calculations, the asymmetry is explained by higher density of states
in the valence band compared to the conduction band.

\section{Discussion}

The influence of magnetic fields on the properties of 2D electrons in HgTe-based
quantum wells is a subject of renewed interest since experimental observation of
the QSHE state in such systems. We have shown that an in-plane magnetic field $B$
profoundly rebuilds the energy spectrum of 2D electrons in these wells owing to
relatively small energy separation of size-quantized subbands. The increase of the
field transforms the insulating QSHE state into a gapless metallic state. This
theoretical conclusion is supported by experimental data on the disappearance
of nonlocal resistance, which takes place around $B=12$ T in 8 nm-wide wells.
The inhomogeneity of the system plays an important role in this phase transition,
since it allows formation of metallic state via percolation at the fields smaller
than those predicted theoretically for ideal wells. By considering two plausible
mechanisms of disorder, smooth variation of well width and electrostatic potential
with reasonable amplitudes, we have calculated the fraction of metallic ($p$) and
insulating ($1-p$) parts of the sample and found that the percolation threshold,
when about half of the 2D plane is in the metallic state ($p \simeq 0.5$), is
realized at 10-12 T, in agreement with our experimental data. The dependence of
the resistance on the magnetic field and gate voltage qualitatively correlates
with the dependence of the insulating fraction on these parameters.

The similarity in behavior of local and nonlocal resistance (Figs.
4 and 5) and persistence of nonlocal resistance up to the field of
percolation suggest that the edge state transport remains essential
up to the transition to metallic state. Indeed, according to the
theory \cite{raichev}, the in-plane magnetic field does not destroy
the edge states. However, the edge states are no longer gapless at
any finite $B$, so that varying the Fermi energy within the bulk gap
by the gate one can always reach the situation when the edge state
transport is absent. Surprisingly, we do not observe such a behavior
in experiment. Another interesting question is how the disorder
affects the edge states and their stability. It is generally
expected that the physics of topological insulator is unaffected by
weak disorder \cite{hasan, qi, moore, moore2, kane, bernevig,
bernevig2}. In the presence of in-plane magnetic field, the
influence of disorder on the edge state transport should be
significant, since we observe a monotonic decrease of nonlocal
resistance which becomes stronger as the field approaches the point
of phase transition. This behavior could be explained by partial
shunting of edge channels by the bulk transport viewed as hopping of
electrons between metallic clusters. Indeed, we observe a decrease
of both local and nonlocal resistance with temperature (Fig. 5)
which can be attributed to increasing probability of such hopping.
Another possible mechanism of decrease of nonlocal resistance with
increasing $B$ is based purely on the edge state transport
properties as described below. The metallic clusters formed at the
sample edges can act as additional leads since an electron entering
a cluster from 1D channel is thermalized by dissipation. The
presence of additional leads along the path between the voltage
probes does not affect the resistance if these leads are assumed to
be point-like. However, since the metallic clusters have finite
sizes expanding with increasing $B$, the effective distance $L$,
which is formed as a sum of all consecutive segments of edge
channels connecting the metallic clusters between the voltage
probes, decreases. According to Eq. (1), the resistance decreases as
well, because at a shorter distance a higher extent of ballisticity
is reached.

In conclusion, we have studied a phase transition of 2D electron system in
HgTe quantum wells from the insulating QSHE state to a metallic state which
is described as a gapless state according to theory. The transition is caused
by application of a magnetic field parallel to the 2D plane. We have emphasized
the crucial role of disorder in our samples and discussed the influence of the
disorder both on the phase transition and on the edge state transport below
the transition field. Whereas our experiments open an interesting possibility
for investigating phase transitions in two dimensions, more experimental and
theoretical work is required to understand the behavior of 2D electron system
in such complex objects as disordered HgTe quantum wells.\\

A financial support of this work by FAPESP, CNPq (Brazilian
agencies), RFBI (N 12-02-00054 and N 11-02-12142-ofi-m) and RAS
programs "Fundamental researches in nanotechnology and
nanomaterials" and "Condensed matter quantum physics" is
acknowledged.

\end{document}